%% file: main.tex
\documentclass{ifacconf}

\usepackage{natbib}        
\usepackage{graphicx}
\usepackage{amsmath}
\usepackage{amssymb}
\usepackage{tikz,pgfplots}
\pgfplotsset{compat=newest}
\usetikzlibrary{patterns}
\usepackage{algorithm}
\usepackage{algorithmic}

\DeclareMathOperator*{\Conv}{Conv}
\DeclareMathOperator*{\Ver}{Ver}

\newtheorem{proposition}{Proposition}
\newtheorem{theorem}{Theorem}
\newtheorem{definition}{Definition}
\newtheorem{lemma}{Lemma}
\newtheorem{corollary}{Corollary}
\newtheorem{remark}{Remark}
\newtheorem{assumption}{Assumption}

\begin{document}
\begin{frontmatter}

\title{A ``Safe Kernel'' Approach for Resilient Multi-Dimensional Consensus} 

\author[First]{Jiaqi Yan} 
\author[Second]{Yilin Mo} 
\author[First]{Xiuxian Li}
\author[First]{Changyun Wen}

\address[First]{School of Electrical and Electronic Engineering, Nanyang Technological University, Singapore. (e-mails: jyan004@e.ntu.edu.sg, xiuxianli@ntu.edu.sg, ecywen@ntu.edu.sg)}
\address[Second]{Department of Automation and BNRist, Tsinghua University, Beijing, China. (e-mail: ylmo@tsinghua.edu.cn)}

\begin{abstract}                
This paper considers the resilient multi-dimensional consensus problem in networked systems, where some of the agents might be malicious (or faulty). 
We propose a multi-dimensional consensus algorithm, where at each time step each healthy agent computes a ``safe kernel'' based on the information from its neighbors, and modifies its own state towards a point inside the kernel.
Assuming that the number of malicious agents is locally (or globally) upper bounded, sufficient conditions on the network topology are presented to guarantee that the benign agents exponentially reach an agreement within the convex hull of their initial states, regardless of the actions of the misbehaving ones. It is also revealed that the graph connectivity and robustness required to achieve the resilient consensus increases linearly with respect to the dimension of the agents' state, indicating the existence of a trade-off between the low communication cost and system security. Numerical examples are provided in the end to validate the theoretical results.  
\end{abstract}


\end{frontmatter}

\section{Introduction}
Recent advances in signal processing and cooperative control have led to growing research interests in networked systems. One of the most important focuses in such systems is the average consensus problem. Given a set of autonomous agents (such as sensors, vehicles, \textit{etc.}), this problem seeks for a distributed protocol that the agents can utilize to reach a common decision/agreement on the average of their initial opinions (see \cite{lynch1996distributed,olfati2007consensus}).

In the past decades, considerable attention has been paid to the development of distributed consensus algorithms (\cite{olfati2007consensus,ren2007information,wei2012distributed}). The existing protocols, although effective in solving the problems under mild conditions, are normally based on the hypothesis that every computing agent is trustworthy and cooperate to follow the algorithms throughout the execution. Nevertheless, as the scale of the network increases, it becomes more difficult to secure every agent. One primary reason is that the widely-adopted communication infrastructures in the distributed framework make it much vulnerable to external adversaries (\cite{mo2012cyber}). Especially, malicious attackers can degrade the algorithm performance by manipulating the transmitted data on communication lines. On the other hand, some agents may not be willing to follow the given rules if they weigh their private interests more than the public ones. They might send out well-designed signals to manipulate the achieved solution for their benefits. It is possible that the misbehaving agents can dictate the final consensus value, or the network may fail to reach an agreement.

It is noted that the consensus problem has been widely applied to the safety-critical systems, such as transportation (\cite{ren2007information,raffard2004distributed}), power grids (\cite{kar2014distributed,kekatos2013distributed}). Since the system failures would cause irreparable harm to economy, environment, and even public health, security and resilience are becoming priory considerations when designing the algorithms (\cite{pasqualetti2012consensus,cardenas2008research}). In recent years, the secure protocols of reaching average consensus in the presence of faulty or misbehaving agents have been widely studied (\cite{dolev1986reaching,leblanc2013resilient,vaidya2012iterative}). 
For example, \cite{dolev1986reaching} consider the approximate consensus problem, where the approximate, rather than exact, agreement is desired in the presence of malicious agents. They consider only complete networks. In order to overrule the effects of malicious nodes, an updating strategy, namely, Mean-Subsequence Reduced \textit{(MSR)} algorithm, is proposed: each normal agent is required to discard the most extreme values in its neighborhood and updates the state based on the remaining values at any time. In a recent work, \cite{leblanc2013resilient} generalize \textit{MSR} to the Weighted Mean-Subsequence-Reduced \textit{(W-MSR)} algorithm. Instead of complete networks, they attempt to analyze this algorithm in more general topologies. A novel property named network robustness is introduced, which characterizes the resilience properties of \textit{W-MSR} in terms of the graph structure. These algorithms, under certain conditions, ensure the agreement within the range of initial values of the normal nodes, even in an adversarial environment.

However, most of the research on resilient consensus assumes that the agents' states are scalar variables, producing crucial limitations in various practical applications, such as vehicle formation control on a 2D-plane.  A naive way to generalize the results on scalar system to multi-dimensional system is to apply \textit{MSR} or \textit{W-MSR} to each entry of the state vectors. The region that the benign agents converge to can be immediately identified as a multi-dimensional ``box'' limited by the minimum and maximum value of their initial states in every dimension. However, is it possible to design a resilient consensus algorithm that provides more accurate convergence results? 

As proved by \cite{su2015fault}, in the presence of the misbehaving agents, it is impossible for any distributed rule to reach the exact average of the initial states of all benign agents. As a compromise, in this paper, we aim to design a multi-dimensional consensus algorithm that converges to a convex combination of these states. In the developed algorithm, each benign agent creates a ``safe kernel'' and modifies its state towards a point inside the kernel. Under certain conditions on network topology, we prove that the proposed strategy guarantees the benign agents of reaching an agreement within the convex hull of their initial values, which improves the accuracy of that by simply applying the existing algorithms to each dimension. It is also noted that the ``safe kernel'' technique can be further extended to other consensus-based problems (\textit{e.g.}, distributed optimization, distributed estimation). Therefore, our work acts as leverage in handling misfunctioning component in multi-dimensional spaces.


\textit{Notations:} For a vector $a$, $a_i$ denotes its $i$-th component. For set $\mathcal{S}\subset \mathbb R^d$, $\Conv(\mathcal{S})$ denotes its convex hull, namely the set of all convex combinations of the points in $\mathcal{S}$. 

\section{Preliminaries}

We start by introducing some technical preliminaries on the graph theory, which would be applied in our further analysis.

Consider the network $\mathcal{G}=\{\mathcal{V},\mathcal{E}\}$, where $\mathcal{V}$ is the set of agents, and $\mathcal{E}\subset \mathcal{V}\times\mathcal{V}$ is the set of edges. An edge between agent $i$ and $j$ is denoted by $e_{ij}\in \mathcal{E}$, indicating these two agents can communicate directly with each other. We define the neighborhood of an agent $i\in \mathcal{V}$ as $$\mathcal{N}_i=\{j\in \mathcal{V}|e_{ij}\in \mathcal{E}\}.$$
Some definitions on the robustness of graph are discussed below (\cite{zhang2015notion}):
\begin{definition}($r$-robust network):
	A network $\mathcal{G}=\{\mathcal{V},\mathcal{E}\}$ is said to be $r$-robust, if for any pair of disjoint and nonempty subsets $\mathcal{V}_1, \mathcal{V}_2\subsetneq\mathcal{V}$, at least one of the following statements hold:
	\begin{enumerate}
		\item There exists more than one agent in $\mathcal{V}_1$, such that it has at least $r$ neighbors outside $\mathcal{V}_1$;
		\item There exists more than one agent in $\mathcal{V}_2$, such that it has at least $r$ neighbors outside $\mathcal{V}_2$.
	\end{enumerate}
\end{definition}
\begin{definition}(($r,s$)-robust network):
	A network $\mathcal{G}=\{\mathcal{V},\mathcal{E}\}$ is said to be ($r,s$)-robust, if for any pair of disjoint and nonempty subsets $\mathcal{V}_1, \mathcal{V}_2\subsetneq\mathcal{V}$, at least one of the following statements hold:
	\begin{enumerate}
		\item Any agent in $\mathcal{V}_1$ has at least $r$ neighbors outside $\mathcal{V}_1$;
		\item Any agent in $\mathcal{V}_2$ has at least $r$ neighbors outside $\mathcal{V}_2$; 
		\item There are no less than $s$ agents in $\mathcal{V}_1\cup\mathcal{V}_2$, such that each of them has at least $r$ neighbors outside the set it belongs to ($\mathcal{V}_1$ or $\mathcal{V}_2$).
	\end{enumerate}
\end{definition}
Intuitively, the definitions of network robustness claim that for any two disjoint and nonempty subsets of agents, there are ``many'' agents within those sets that have a sufficient number of neighbors outsides. As we will see, the robust graph plays an important role in our analysis of achieving a resilient agreement. 

\section{Problem Formulation}\label{sec:form}
Consider the network modeled by an undirected and connected graph $\mathcal{G}=\{\mathcal{V},\mathcal{E}\}$, where $\mathcal{V}=\{1,2,...,N\}$. At any time $k\geq 0$, let $x^i(k)\in\mathbb{R}^d$ denote the current state of agent $i$. The agents are said to reach a (distributed) consensus if and only if there exists a constant $\tilde x$, such that $\lim_{k\rightarrow \infty}x^i(k)=\tilde{x}$ holds for every agent $i$. In particular, if $\tilde{x}=1/N\sum_{i=1}^{N}x^i(0)$, an average consensus is achieved. 

Many practical applications fit into the framework of average consensus (see \cite{ren2007information,xiao2007distributed}). While various strategies have been developed to facilitate it, the linear algorithms have attracted much attention due to their simplicity and ease of implementation. In such strategies, every agent $i\in\mathcal{V}$ is initialized with $x^i(0)$. At each time $k$, it receives information from all of its neighbors, and updates its own state according to the following equation: 
\begin{equation}\label{eqn:updatewithoutbenign}
x^i(k+1) = a_i^i(k)x^i(k)+\sum_{j\in\mathcal{N}_i}a_j^i(k)x^j(k),
\end{equation}
The new state will then be broadcasted to its neighbors preparing for the next updating stage. The conditions under different scenarios to ensure the achievement of average consensus have been investigated widely in the literatures (see \cite{nedic2010constrained,olfati2007consensus}), the details of which are omitted here due to the space limitation.  

We should note that, an implicit assumption for the effectiveness of this approach, and other distributed algorithms as well, is that all agents are reliable throughout the execution, and cooperate to achieve the desired value. However, as the number of local agents increases, certain concerns arise that might make this assumption to be violated. As discussed before, its strong dependence on the communication infrastructures creates lots of vulnerabilities for cyber attacks, where the transmitted information might be manipulated by external adversaries. Additionally, ``non-participant'' agent may exist, who deviates from the normal update rule and sends out self-designed information for its own benefits. Clearly, such illegal behaviors would degrade the performance of distributed protocols: they can either prevent the benign agents from reaching a consensus, or manipulate the final agreement to be false. 

The security concerns lead to the study of resilient consensus protocols. By saying ``resilient'', we hope to achieve the following objectives, regardless of the choice of initial states and even in the adversarial environment:
\begin{enumerate}
	\item\emph{Agreement:} As $k$ goes to infinity, it is held that $x^i(k) = \bar{x}$ with some $\bar{x}\in\mathbb{R}^d$, for any benign agent $i$; 
	\item\emph{Validity:} At any time and for any benign agent, its state remains in the convex hull of all benign agents' initial values.
\end{enumerate}

We elucidate these conditions as below. Firstly, the states of the benign agents should converge to the same constant value even in the presence of misbehaving ones. In addition, they are not allowed to leave the convex hull of their initial states throughout the procedure. That is, they should avoid being influenced by the misbehaviors too much. It is observed that if $1$D problem is considered, then the validity condition would be degraded to that ``\textit{the state of any benign agent always remains in the interval forming by the minimum and maximum of their initial states}''. There has been much work proved to be effective in this simple case (\textit{e.g.},  \textit{MSR} proposed by \cite{dolev1986reaching} and \textit{W-MSR} proposed by \cite{leblanc2013resilient}). However, few research efforts have been devoted to the more general multi-dimensional systems. 

\begin{figure}[!htbp]
	\centering
	\input{tikz/boxvsconvexhull.tikz}
	\caption{A $2$D illustration with agents marked with circles. The location of the node indicates its initial value. With the direct application of existing algorithms to each dimension, the final agreement is ensured to be within the rectangle represented by oblique lines. However, a better solution satisfying the validity condition of converging to the solid triangle is expected. }
	\label{Fig: boxvsconvexhull}
\end{figure}
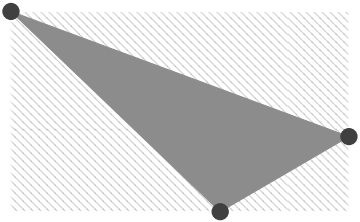

A naive way to tackle this problem is by simply applying the existing scalar protocols to each component of the state vectors. Nevertheless, the region that the benign agents converge to can only be guaranteed as a multi-dimensional ``box''  limited by the minimum and maximum value of their initial states at every dimension, and thus the validity condition fails to be ensured in this manner. To see this, we present a $2$-dimensional illustration in Fig. \ref{Fig: boxvsconvexhull}, indicating this naive algorithm cannot guarantee the convergence to a point inside the convex hull of initial states. Therefore, this paper intends to address this problem and come up with a method satisfying both Conditions 1) and 2).  

\subsection{Attack model}
We define $\mathcal{F}$ as the set of malicious/faulty agents. Any agent $i\in \mathcal{F}$ could either be the adversarial one with the value being manipulated by the attacker,  or the non-participant agent who does not follow the standard updating rule. We also denote $\mathcal{B}$ as the collection of benign agents who will always follow the prescribed updating strategy. It is clear that $\mathcal{B}\cap\mathcal{F}=\varnothing$ and $\mathcal{B}\cup\mathcal{F}=\mathcal{V}$. 

The faulty nodes could be characterized by the scope of threats:
\begin{enumerate}
	\item ($F$-total attack model) There are at most $F$ misbehaving agents in the network. That is, $|\mathcal{F}|\leq  F$.
	\item ($F$-local attack model) There are at most $F$ misbehaving agents in the neighborhood of any benign agent. That is, $|\mathcal{F}\cap\mathcal{N}_i|\leq F$, for any agent $i\in\mathcal{B}$.
\end{enumerate}
It is easy to conclude that  $F$-total attack model is a special case of $F$-local one.


Note that we do not pose any restrictions on the transmitted information of agent $i\in\mathcal{F}$, \textit{i.e.}, the malicious agents are allowed to send out arbitrary data to their neighbors. Furthermore, they could collude among themselves to decide on the deceptive values to be communicated. 

\section{A Resilient Multi-dimensional Consensus Strategy}\label{sec:alg}
In this section, we provide a resilient consensus algorithm. To simplify notations, we have the following definitions:
\begin{definition}
	Consider a set $\mathcal{A}\subset \mathbb R^d$ with cardinality $m$\footnote{To be more precise, $\mathcal A$ should be defined as a multi-set since we allow duplicate elements in the set, e.g., the states of $m$ agents shall be counted as $m$ points even if some of them may be identical.}. Let $\mathcal{S}(\mathcal{A},n)$ be the set of all its subset with cardinality $m-n$. 
\end{definition}

It is clear that the set $\mathcal{S}(\mathcal{A},n)$ contains $\binom{m}{n}$ elements, and each of them is associated with a convex hull. The intersection of all these convex hulls plays a crucial role in our algorithm, which is defined as follows: 
\begin{definition}\label{def: intersection}
	Consider the set $\mathcal{A}\subset\mathbb R^d$ with cardinality $m$. We define $\varPsi (\mathcal{A}, n)$ as 
	\begin{align}
	\varPsi(\mathcal A,n) \triangleq \bigcap_{S\in \mathcal S(\mathcal A,n)} \Conv(S).
	\end{align}
\end{definition}

Given the $F$-total/ $F$-local attack model introduced before, the designed algorithm is formally presented as follows. Each agent $j\in \mathcal{V}$ is initialized with a starting state $x^j(0)\in \mathbb{R}^d$. At any time $k> 0$, every benign agent $i\in\mathcal{B}$ updates as outlined in Algorithm \ref{alg:resilient}.
\begin{algorithm}
	1:\: Receive the states from all neighboring agents $j\in\mathcal{N}_i$, and collect these values in $\mathcal{X}^i(k)$.\\
	2:\: Define $\mathcal{R}^i(k)\triangleq\varPsi (\mathcal{X}^i(k), F)$, and denote the vertices of this set to be $\Ver(\mathcal{R}^i(k))$. Agent $i$ updates its local state as:
	\begin{equation}\label{eqn:updatewithattack}
	\begin{split}
	x^i(k+1) = a_i^i(k)x^i(k)+\sum_{\bar{x}^j(k)\in\Ver(\mathcal{R}^i(k))}a_j^i(k)\bar{x}^j(k),
	\end{split}
	\end{equation} \\
	satisfying that each weight is lower bounded by some $\alpha>0$, and $a_i^i(k)+\sum_{\bar{x}^j(k)\in\Ver(\mathcal{R}^i(k))}a_j^i(k)=1$.\\
	3:\: Transmit updated state $x^i(k+1)$ to all neighbors $j\in\mathcal{N}_i$. \\
	\caption{Resilient consensus algorithm}
	\label{alg:resilient}
\end{algorithm}

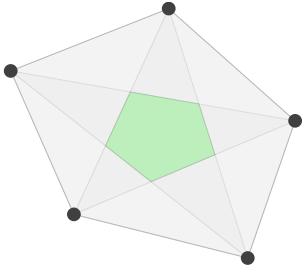
\begin{figure}
	\centering
	\input{tikz/safekernel.tikz}
	\caption{A $2$D illustration of ``safe kernel''. Suppose agent $i\in\mathcal{B}$ has $5$ neighbors and each of their states is represented by the location of a circle. Let $F=1$. The green region denotes $\mathcal{R}^i(k)=\varPhi(\mathcal{X}^i(k), 1)$, namely the ``safe kernel''. }
	\label{fig:safekernel}
\end{figure}

\begin{remark}
	We can interpret $\mathcal{R}^i(k)$ as the ``safe kernel'' (illustrated in Fig. \ref{fig:safekernel}), which is guaranteed to be within the convex hull forming by only benign ones, as we will prove later. Intuitively, at any time, the healthy agent computes and moves its state toward a point inside the ``safe kernel''. As a result, the impact of malicious agents on the benign ones are limited. The proposed protocol can be implemented in a distributive fashion, as every fault-free agent is only required to access the local information, with no need to have any knowledge of the network topology or impose extra communication among agents. 
\end{remark}

\section{Algorithm Analysis}\label{sec:analysis}
This section is devoted to proving the effectiveness of Algorithm \ref{alg:resilient}. In this paper, we impose the following assumption on the network topology:
\begin{assumption}\label{assump:connect}
	For any $i\in\mathcal{V}$, it is held that $|\mathcal N_i|\geq (d+1)F+1$.
\end{assumption}
\subsection{Realizability}
Before analyzing the resiliency of the proposed algorithm, we need to prove its realizability. First, we show that $\mathcal R^i(k)$ is non-empty. To this end, we shall begin with the introduction of Helly's theorem, which is a basic result of convexity theory and key supporting technique of this paper.
\begin{theorem}[Helly's theorem](\cite{Danzer1963Helly})\label{thm:Helly}
	Let $X_1,\cdots, X_p$ be a finite collection of convex subsets in $\mathbb{R}^d$, with $p > d$. If the intersection of every $d + 1$ of these sets is nonempty, then the whole collection has a nonempty intersection. That is,
	$$\bigcap _{j=1}^{p}X_{j}\neq \varnothing. $$
\end{theorem}

A direct result of Helly's theorem is as follows:
\begin{corollary}\label{lm:existence}
	Let $\mathcal{A}$ be a set with cardinality $m$ in $\mathbb{R}^d$. If $m\geq n(d+1)+1$, then for any $n\leq m$, the following relation holds $$\varPsi (\mathcal{A}, n) \neq \varnothing.$$
\end{corollary}
%

By Corollary \ref{lm:existence}, we note that Assumption \ref{assump:connect} guarantees that $\mathcal{R}^i(k) \neq \varnothing$ for any $i\in\mathcal{B}$ at any time. 

Next we need to show the existence of $a_j^i(k)$. Notice that the number of vertices of $\mathcal R^i(k)=\varPsi (\mathcal{X}^i(k), F)$ is upper bounded\footnote{To see this, notice that each convex hull in $\mathcal S(\mathcal X^i(k),F)$ is a polytope limited with $|\mathcal X^i(k)|-F$ number of vertices and its number of facets is bounded due to the Upper Bound Theorem (\cite{ziegler2012lectures}). As each vertex of $\mathcal{R}^i(k)$ is an intersection of at least $d$ of these facets, we know that its number is upper bounded.}, so that the lower bound of the weights $\alpha$ exists. 
Therefore, one concludes that the proposed strategy is realizable.

\subsection{Resiliency}
Before proceeding to the main results, we shall first present some preliminary conclusions regarding $\varPsi (\mathcal{A}, n)$ [cf. Definition \ref{def: intersection}]:

\begin{proposition}\label{pro:subset}
	Consider two collections of sets $\left\{A_i\right\}_{i\in I}$ and $\left\{B_j\right\}_{j\in \mathcal J}$. If for any $j\in \mathcal J$, there exists an $i\in \mathcal I$, such that $B_j\supseteq A_i$, then
	\begin{align*}
	\bigcap _{j\in\mathcal J} B_j \supseteq\bigcap_{i\in\mathcal I}A_i.
	\end{align*}
\end{proposition}
\begin{lemma}\label{lm: subset}
	Consider any set $\mathcal{A}_1$ with cardinality $m_1$ and $\mathcal{A}_2$ with cardinality $m_2$. If $\mathcal{A}_1 \subset \mathcal{A}_2$, then for any $n\leq m_1$, the following statement holds: $$\varPsi (\mathcal{A}_1, n) \subset \varPsi (\mathcal{A}_2, n).$$
\end{lemma}

\begin{lemma}\label{lm:y^i}
	Let $\mathcal{A}$ be a set with cardinality $m$ in $\mathbb{R}^d$. Supposing that $m\geq (d+1)n+1$, the following relations hold for any $n\leq m$ and any $p\in\{1,2,..,d\}$:
	\begin{enumerate}
		\item If no more than $n$ elements of $\mathcal{A}$ has its $p^{th}$ entry greater than $\varepsilon$, then for any $y \in\varPsi (\mathcal{A}, n)$, it is held that $y_p\leq \varepsilon$;
		\item If no more than $n$ elements of $\mathcal{A}$ has its $p^{th}$ entry less than $\varepsilon$, then for any $y \in\varPsi (\mathcal{A}, n)$, it is held that $y_p\geq \varepsilon$.
	\end{enumerate}
\end{lemma}

Now we are ready to provide our main results. For simplicity, we denote the convex hull of the states of all benign agents at time $k$ as $\varOmega(k)$. The following theorem presents the non-expansion property of $\varOmega(k)$: 
\begin{theorem}[Validity]\label{thm:safety}
	Consider the network $\mathcal{G}(\mathcal{V}, \mathcal{E})$. With Algorithm \ref{alg:resilient}, the following relation holds for any $k\geq 0$:
	\begin{equation}
	\varOmega(k+1)\subset \varOmega(k),
	\end{equation}
	under either $F$-local or $F$-total attack model.
\end{theorem}
%
%

Theorem \ref{thm:safety} indicates that the proposed algorithm guarantees the validity condition of resilient consensus. That is, the healthy agents would never be out of the convex hull of their initial values, despite the influence of the misbehaving agents. In what follows, we will provide sufficient conditions on network topology, under which the agreement condition will also be satisfied. Due to the space limitation, the proof of these theorems are omitted. 
\begin{theorem}[Agreement: $F$-local]\label{thm:consensus_local}
	Consider the network $\mathcal{G}(\mathcal{V}, \mathcal{E})$. Suppose the misbehaving agents follow an $F$-local attack model. If the network is with $((d+1)F+1)$-robustness, then with Algorithm \ref{alg:resilient}, all the benign agents are guaranteed to achieve consensus exponentially, regardless of the actions of misbehaving agents. 
\end{theorem}

The next theorem elaborates a different condition for the proposed algorithm to succeed under $F$-total threats: 
\begin{theorem}[Agreement: $F$-total]\label{thm:consensus_total}
	Consider the network $\mathcal{G}(\mathcal{V}, \mathcal{E})$. Suppose the misbehaving agents follow an $F$-total attack model. If the network is with $(dF+1,F+1)$-robustness, then with Algorithm \ref{alg:resilient}, all the benign agents are guaranteed to achieve consensus exponentially, regardless of the actions of misbehaving agents. 
\end{theorem}
%
%
\begin{remark}
	By definitions, it is easy to see that a $((d+1)F+1)$-robust graph is $(dF+1,F+1)$-robust as well, but not vice versa. That is to say, the network which is able to tolerate $F$-local attacks could also survive the $F$-total ones, while the converse is not true. This observation is consistent with the fact that the $F$-globally bounded threats are special versions of locally bounded ones.
\end{remark}

Based on the above results, one obtains immediately that the proposed algorithm facilitates the resilient consensus. We formally state it in the next theorem:
\begin{theorem}\label{thm:converge}
	Consider the network $\mathcal{G}(\mathcal{V}, \mathcal{E})$. Suppose the network satisfies one of the following conditions:\\
	1) under $F$-local attack model, and is $((d+1)F+1)$-robust,\\
	2) under $F$-total attack model, and is $(dF+1,F+1)$-robust.\\
	With Algorithm \ref{alg:resilient}, all the benign agents finally achieve a consensus within the convex hull of the initial states of benign agents, regardless of
	the actions of misbehaving ones. That is, as $k\to\infty$,
	\begin{equation}
	x^i(k)=x^j(k) = \hat{x} \quad\mbox{ for any } i, j\in\mathcal{B},
	\end{equation}
	where
	$\hat{x}\in \varOmega(0).$ 
\end{theorem}

\begin{remark}
	Since the convergence of proposed algorithm does not depend on the actions of misbehaving agents, it works effectively even in the worst-case scenario, where the misbehaving agents could have full knowledge of graph topology, updating rules, \textit{etc}, and could be able to send different data to different neighbors. 
\end{remark}

Theorem \ref{thm:converge} indicates that under certain requirements on network topology, Algorithm \ref{alg:resilient} guarantees that all benign agents reach an agreement on a weighted average of their initial states, \textit{i.e.}, $\hat{x}=\sum_{i\in\mathcal{B}}\gamma_{i} x_i(0)$ with $\gamma _{i}\geq 0$ and $\sum _{i\in\mathcal{B}}\gamma_{i}=1$. As proved by \cite{su2015fault}, if $\mathcal{F}$ is nonempty, it is impossible for any distributed rule to achieve the exact average of these states. Therefore, our algorithm is effective in the sense that a suboptimal result is achieved. It protects the states of the benign agents from being driven to arbitrary values, and thus could withstand the compromise of partial agents while providing a desired level of security. 

\section{Numerical Example}\label{sec:casestudy}
\begin{figure}[!htbp]
	\centering
	\input{tikz/topology.tikz}
	\caption{Communication network.}
	\label{Fig:network}
\end{figure}
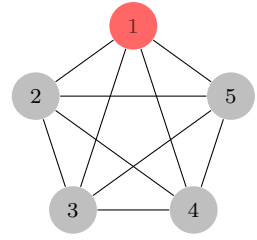

In this section, we provide numerical examples to verify the theoretical results established in the previous sections. In the example, the communication network is given by Fig. \ref{Fig:network}, in which the node set is $\mathcal{V} = \{1,2, . . .,5\}$. It is verified that the graph is $(3,2)$-robust. Suppose that agent $1$ is compromised. It intends to prevent others from reaching a correct consensus by violating the rule in Algorithm \ref{alg:resilient} and setting its states as $x^1_1(k) = 1.5*\sin(k/5)$ and $x^1_2(k) = k/25+1$ at any time $k>0$.  On the other hand, the benign agents are initialized with $x^2(0)=(1,2), x^3(0)=(2,0), x^4(0)=(1,3), x^5(0)=(2,4)$, and always follow \eqref{eqn:updatewithattack} as updates. For simplicity, let their updating weights be $ a_j^i(k)= 1/(|\Ver(\mathcal{R}^i(k))|+1)$ for each $j \in \Ver(\mathcal{R}^i(k)) \cup \{i\}$. 

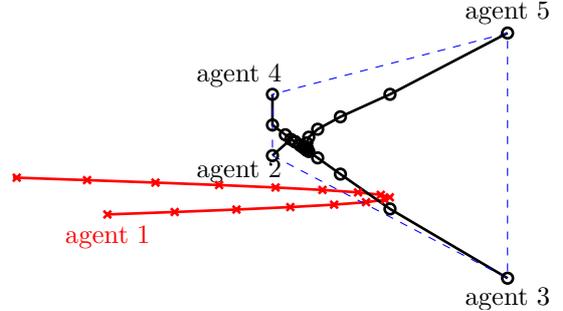
\begin{figure}[!htbp]
	\centering
	\input{tikz/test.tikz}
	\caption{The trajectory of local states under Algorithm \ref{alg:resilient}, where the area surrounded by the dashed lines is the convex hull of the initial states of benign agents.}
	\label{fig:trajectory}
\end{figure}
%

We test the performance of Algorithm \ref{alg:resilient} in Fig. \ref{fig:trajectory}. The result shows that the states of benign agents are always guaranteed within the convex hull of their initial states and they finally achieve a common value, which validates Theorem \ref{thm:consensus_total}. That is, the network could tolerate a single misbehaving node in this $2$-dimensional problem. Since the malicious agent is unable to affect the final agreement too much, our protocol helps to improve the system security.


\section{Conclusion}\label{sec:conclusion}
Due to its wide applications, the problem of average consensus attracts much research interest in recent years. In this paper, we are interested in the achievement of average consensus under malicious agents in the multi-dimensional spaces. We propose a resilient distributed algorithm. Under certain network topology, the designed protocol is proved to guarantee that all benign agents exponentially reach an agreement within the convex hull of their initial states, regardless of the actions of faulty ones. 

The future work involves the design of a more effective algorithm in the scenario where the network topology fails to meet the sufficient conditions. Furthermore, the theoretical analysis of the accomodation of ``safe kernel'' techinique to other problem settings is also a possible research direction.

\bibliography{main}             

\end{document}

%% file: tikz/boxvsconvexhull.tikz
\begin{tikzpicture}

\begin{axis}[%
width=2.0in,
height=1.15in,
color= white,
scale only axis,
xtick=\empty,
xmin=-0.2,
xmax=2.2,
ytick=\empty,
ymin=-0.2,
ymax=4.2,
axis lines=middle, 
]
\draw[pattern=north west lines,pattern color=gray!50!black,fill opacity=0.5] (0,0) rectangle (2.1,4);
\draw[color=gray!90!white,fill=gray!90!white,fill opacity=1] (0,4)--(1.3,0)--(2.1,1.5)--(0,4);
\node[shape=circle,inner sep=1.5pt,fill=gray!50!black,fill opacity=1] (1) at (0,4) {$\;$};
\node[shape=circle,inner sep=1.5pt,fill=gray!50!black,fill opacity=1] (2) at (1.3,0) {$\;$};
\node[shape=circle,inner sep=1.5pt,fill=gray!50!black,fill opacity=1] (3) at (2.1,1.5) {$\;$};
\end{axis}
\end{tikzpicture}%

%% file: tikz/safekernel.tikz
\begin{tikzpicture}[fill opacity=0.1]

\begin{axis}[%
width=1.8in,
height=1.4in,
color= white,
scale only axis,
xtick=\empty,
xmin=-0.2,
xmax=2,
ytick=\empty,
ymin=-0.2,
ymax=4.1,
axis lines=middle, 
]
\draw[color=black,fill=gray,fill opacity=0.03] (0,3)--(0.4,0.7)--(1.5,0)--(1,4)--(0,3);
\draw[color=black,fill=gray,fill opacity=0.03] (0,3)--(0.4,0.7)--(1.5,0)--(1.8,2.2)--(0,3);
\draw[color=black,fill=gray,fill opacity=0.03] (0,3)--(0.4,0.7)--(1.8,2.2)--(1,4)--(0,3);
\draw[color=black,fill=gray,fill opacity=0.03] (0,3)--(1.5,0)--(1.8,2.2)--(1,4)--(0,3);
\draw[color=black,fill=gray,fill opacity=0.03] (0.4,0.7)--(1.5,0)--(1.8,2.2)--(1,4)--(0.4,0.7);
\node[shape=circle,inner sep=1.0pt,fill=gray!50!black,fill opacity=1] (1) at (0,3) {$\;$};
\node[shape=circle,inner sep=1.0pt,fill=gray!50!black,fill opacity=1] (2) at (0.4,0.7) {$\;$};
\node[shape=circle,inner sep=1.0pt,fill=gray!50!black,fill opacity=1] (3) at (1.5,0) {$\;$};
\node[shape=circle,inner sep=1.0pt,fill=gray!50!black,fill opacity=1] (4) at (1.8,2.2) {$\;$};
\node[shape=circle,inner sep=1.0pt,fill=gray!50!black,fill opacity=1] (5) at (1,4) {$\;$};
\draw[color=black,fill=green,fill opacity=0.2] (0.6,1.8)--(0.7570,2.6635)--(1.1912,2.4706)--(1.2929,1.6568)--(0.8883,1.2233)--(0.600,1.8);
\end{axis}
\end{tikzpicture}%

%% file: tikz/topology.tikz
\begin{tikzpicture}[scale =0.45] 
\coordinate (O) at (0, 0);
\node[shape=circle,inner sep=4pt,fill=red,fill opacity=0.6] (1) at (0,3) {{\small $1$}};
\node[shape=circle,inner sep=4pt,fill=lightgray] (2) at (90+72:3cm) {{\small $2$}};
\node[shape=circle,inner sep=4pt,fill=lightgray] (3) at (90+2*72:3cm) {{\small $3$}};
\node[shape=circle,inner sep=4pt,fill=lightgray] (4) at (90+3*72:3cm) {{\small $4$}};
\node[shape=circle,inner sep=4pt,fill=lightgray] (5) at (90+4*72:3cm) {{\small $5$}};

\draw (1)--(2)--(3)--(4)--(5)--(1);
\draw (1)--(3)--(5);
\draw (1)--(4);
\draw (5)--(2)--(4);

\end{tikzpicture}

%% file: tikz/test.tikz
%
%
%
\begin{tikzpicture}

\begin{axis}[%
width=2.8in,
height=1.6in,
color= white,
at={(1.011in,0.642in)},
scale only axis,
xtick=\empty,
xmin=-0.1,
xmax=2.2,
ytick=\empty,
ymin=-0.5,
ymax=4.5,
axis lines=middle, 
]
\addplot [color=red,line width=1.0pt, mark = x]
table[row sep=crcr]{%
0.298003996	1.04\\
0.584127513	1.08\\
0.84696371	1.12\\
1.076034136	1.16\\
1.262206477	1.2\\
1.398058629	1.24\\
1.478174595	1.28\\
1.499360405	1.32\\
1.460771446	1.36\\
1.36394614	1.4\\
1.212744606	1.44\\
1.013194771	1.48\\
0.773252058	1.52\\
0.502482225	1.56\\
0.211680012	1.6\\
-0.087561215	1.64\\
-0.383311653	1.68\\
-0.663780665	1.72\\
-0.917786836	1.76\\
-1.135203743	1.8\\
};

\addplot [color=black, line width=1.0pt, mark = o]
  table[row sep=crcr]{%
1	2\\
1.077954904	2.266026988\\
1.095655031	2.227752538\\
1.116494502	2.170890633\\
1.137305683	2.113862968\\
1.14585598	2.090403128\\
1.144576534	2.123117207\\
1.144576534	2.123117207\\
1.144576534	2.123117207\\
1.144576534	2.123117207\\
1.144576534	2.123117207\\
1.144576534	2.123117207\\
1.144576534	2.123117207\\
1.144576534	2.123117207\\
1.144576534	2.123117207\\
1.144576534	2.123117207\\
1.144576534	2.123117207\\
1.144576534	2.123117207\\
1.144576534	2.123117207\\
1.144576534	2.123117207\\
};

\addplot [color=black, line width=1.0pt, mark = o]
table[row sep=crcr]{%
2	0\\
1.500007456	1.130450734\\
1.28898117	1.698238904\\
1.192318099	1.962995732\\
1.1544063	2.06694319\\
1.145855983	2.090403121\\
1.144576534	2.123117207\\
1.144576534	2.123117207\\
1.144576534	2.123117207\\
1.144576534	2.123117207\\
1.144576534	2.123117207\\
1.144576534	2.123117207\\
1.144576534	2.123117207\\
1.144576534	2.123117207\\
1.144576534	2.123117207\\
1.144576534	2.123117207\\
1.144576534	2.123117207\\
1.144576534	2.123117207\\
1.144576534	2.123117207\\
1.144576534	2.123117207\\
};

\addplot [color=black, line width=1.0pt, mark = o]
table[row sep=crcr]{%
1	3\\
1.000000017	2.499999951\\
1.054731352	2.340606541\\
1.096006903	2.227146771\\
1.127061604	2.141971019\\
1.14073393	2.104457191\\
1.144576534	2.123117207\\
1.144576534	2.123117207\\
1.144576534	2.123117207\\
1.144576534	2.123117207\\
1.144576534	2.123117207\\
1.144576534	2.123117207\\
1.144576534	2.123117207\\
1.144576534	2.123117207\\
1.144576534	2.123117207\\
1.144576534	2.123117207\\
1.144576534	2.123117207\\
1.144576534	2.123117207\\
1.144576534	2.123117207\\
1.144576534	2.123117207\\
};

\addplot [color=black, line width=1.0pt, mark = o]
table[row sep=crcr]{%
2	4\\
1.499999997	2.999999992\\
1.288977448	2.633013484\\
1.192315874	2.430381373\\
1.154404603	2.300621326\\
1.145860241	2.207205388\\
1.144576534	2.123117207\\
1.144576534	2.123117207\\
1.144576534	2.123117207\\
1.144576534	2.123117207\\
1.144576534	2.123117207\\
1.144576534	2.123117207\\
1.144576534	2.123117207\\
1.144576534	2.123117207\\
1.144576534	2.123117207\\
1.144576534	2.123117207\\
1.144576534	2.123117207\\
1.144576534	2.123117207\\
1.144576534	2.123117207\\
1.144576534	2.123117207\\
};

\draw[dashed, color=blue] (1,2)--(2,0)--(2,4)--(1,3)--(1,2);
\node[below, align=left,color=red] at (0.3,1.0) {agent $1$};
\node[left, align=left,color=black] at (1.08,1.75) {agent $2$};
\node[below, align=left,color=black] at (2,0) {agent $3$};
\node[left, align=left,color=black] at (1.08,3.3) {agent $4$};
\node[above, align=left,color=black] at (2,4) {agent $5$};
\end{axis}
\end{tikzpicture}%

%% file: main.bbl
\begin{thebibliography}{19}
\providecommand{\natexlab}[1]{#1}
\providecommand{\url}[1]{\texttt{#1}}
\providecommand{\urlprefix}{URL }
\expandafter\ifx\csname urlstyle\endcsname\relax
  \providecommand{\doi}[1]{doi:\discretionary{}{}{}#1}\else
  \providecommand{\doi}{doi:\discretionary{}{}{}\begingroup
  \urlstyle{rm}\Url}\fi

\bibitem[{C{\'a}rdenas et~al.(2008)C{\'a}rdenas, Amin, and
  Sastry}]{cardenas2008research}
C{\'a}rdenas, A.A., Amin, S., and Sastry, S. (2008).
\newblock Research challenges for the security of control systems.
\newblock In \emph{HotSec}.

\bibitem[{Danzer et~al.(1963)Danzer, Grünbaum, and Klee}]{Danzer1963Helly}
Danzer, L., Grünbaum, B., and Klee, V. (1963).
\newblock Helly's theorem and its relatives.
\newblock \emph{Proceedings of Symposia in Pure Mathematics}, 101--180.

\bibitem[{Dolev et~al.(1986)Dolev, Lynch, Pinter, Stark, and
  Weihl}]{dolev1986reaching}
Dolev, D., Lynch, N.A., Pinter, S.S., Stark, E.W., and Weihl, W.E. (1986).
\newblock Reaching approximate agreement in the presence of faults.
\newblock \emph{Journal of the ACM (JACM)}, 33(3), 499--516.

\bibitem[{Kar et~al.(2014)Kar, Hug, Mohammadi, and Moura}]{kar2014distributed}
Kar, S., Hug, G., Mohammadi, J., and Moura, J.M. (2014).
\newblock Distributed state estimation and energy management in smart grids: A
  consensus$+$innovations approach.
\newblock \emph{IEEE Journal of selected topics in signal processing}, 8(6),
  1022--1038.

\bibitem[{Kekatos and Giannakis(2013)}]{kekatos2013distributed}
Kekatos, V. and Giannakis, G.B. (2013).
\newblock Distributed robust power system state estimation.
\newblock \emph{IEEE Transactions on Power Systems}, 28(2), 1617--1626.

\bibitem[{LeBlanc et~al.(2013)LeBlanc, Zhang, Koutsoukos, and
  Sundaram}]{leblanc2013resilient}
LeBlanc, H.J., Zhang, H., Koutsoukos, X., and Sundaram, S. (2013).
\newblock Resilient asymptotic consensus in robust networks.
\newblock \emph{IEEE Journal on Selected Areas in Communications}, 31(4),
  766--781.

\bibitem[{Lynch(1996)}]{lynch1996distributed}
Lynch, N.A. (1996).
\newblock \emph{Distributed algorithms}.
\newblock Elsevier.

\bibitem[{Mo et~al.(2012)Mo, Kim, Brancik, Dickinson, Lee, Perrig, and
  Sinopoli}]{mo2012cyber}
Mo, Y., Kim, T.H.J., Brancik, K., Dickinson, D., Lee, H., Perrig, A., and
  Sinopoli, B. (2012).
\newblock Cyber--physical security of a smart grid infrastructure.
\newblock \emph{Proceedings of the IEEE}, 100(1), 195--209.

\bibitem[{Nedic et~al.(2010)Nedic, Ozdaglar, and
  Parrilo}]{nedic2010constrained}
Nedic, A., Ozdaglar, A., and Parrilo, P.A. (2010).
\newblock Constrained consensus and optimization in multi-agent networks.
\newblock \emph{IEEE Transactions on Automatic Control}, 55(4), 922--938.

\bibitem[{Olfati-Saber et~al.(2007)Olfati-Saber, Fax, and
  Murray}]{olfati2007consensus}
Olfati-Saber, R., Fax, J.A., and Murray, R.M. (2007).
\newblock Consensus and cooperation in networked multi-agent systems.
\newblock \emph{Proceedings of the IEEE}, 95(1), 215--233.

\bibitem[{Pasqualetti et~al.(2012)Pasqualetti, Bicchi, and
  Bullo}]{pasqualetti2012consensus}
Pasqualetti, F., Bicchi, A., and Bullo, F. (2012).
\newblock Consensus computation in unreliable networks: A system theoretic
  approach.
\newblock \emph{IEEE Transactions on Automatic Control}, 57(1), 90--104.

\bibitem[{Raffard et~al.(2004)Raffard, Tomlin, and
  Boyd}]{raffard2004distributed}
Raffard, R.L., Tomlin, C.J., and Boyd, S.P. (2004).
\newblock Distributed optimization for cooperative agents: Application to
  formation flight.
\newblock In \emph{Decision and Control, 2004. CDC. 43rd IEEE Conference on},
  volume~3, 2453--2459. IEEE.

\bibitem[{Ren et~al.(2007)Ren, Beard, and Atkins}]{ren2007information}
Ren, W., Beard, R.W., and Atkins, E.M. (2007).
\newblock Information consensus in multivehicle cooperative control.
\newblock \emph{IEEE Control systems magazine}, 27(2), 71--82.

\bibitem[{Su and Vaidya(2015)}]{su2015fault}
Su, L. and Vaidya, N.H. (2015).
\newblock Fault-tolerant distributed optimization (part iv): constrained
  optimization with arbitrary directed networks.
\newblock \emph{arXiv preprint arXiv:1511.01821}.

\bibitem[{Vaidya et~al.(2012)Vaidya, Tseng, and Liang}]{vaidya2012iterative}
Vaidya, N.H., Tseng, L., and Liang, G. (2012).
\newblock Iterative approximate byzantine consensus in arbitrary directed
  graphs.
\newblock In \emph{Proceedings of the 2012 ACM symposium on Principles of
  distributed computing}, 365--374. ACM.

\bibitem[{Wei and Ozdaglar(2012)}]{wei2012distributed}
Wei, E. and Ozdaglar, A. (2012).
\newblock Distributed alternating direction method of multipliers.
\newblock In \emph{2012 IEEE 51st IEEE Conference on Decision and Control
  (CDC)}, 5445--5450. IEEE.

\bibitem[{Xiao et~al.(2007)Xiao, Boyd, and Kim}]{xiao2007distributed}
Xiao, L., Boyd, S., and Kim, S.J. (2007).
\newblock Distributed average consensus with least-mean-square deviation.
\newblock \emph{Journal of parallel and distributed computing}, 67(1), 33--46.

\bibitem[{Zhang et~al.(2015)Zhang, Fata, and Sundaram}]{zhang2015notion}
Zhang, H., Fata, E., and Sundaram, S. (2015).
\newblock A notion of robustness in complex networks.
\newblock \emph{IEEE Transactions on Control of Network Systems}, 2(3),
  310--320.

\bibitem[{Ziegler(2012)}]{ziegler2012lectures}
Ziegler, G.M. (2012).
\newblock \emph{Lectures on polytopes}, volume 152.
\newblock Springer Science \& Business Media.

\end{thebibliography}
